\documentclass[sigconf]{acmart}
\renewcommand\footnotetextcopyrightpermission[1]{} 
\usepackage{booktabs} 

\usepackage{braket}
\usepackage{algorithm}

\usepackage{hyperref}
\usepackage{algpseudocode}
\usepackage{multirow}
\usepackage{makecell}
\usepackage{bm}

\definecolor{darkpastelgreen}{rgb}{0.01, 0.75, 0.24}
\renewcommand{\figurename}{Fig.}
\renewcommand{\sectionname}{Section}

\setcopyright{none}
\begin{document}
\title{Towards Optimizations of Quantum Circuit Simulation for Solving Max-Cut Problems with QAOA}
  
\renewcommand{\shorttitle}{SIG Proceedings Paper in LaTeX Format}

\author{Yu-Cheng Lin}
\affiliation{%
  \institution{National Taiwan University}
  \city{Taipei}
  \country{Taiwan}
}
\email{r11922015@csie.ntu.edu.tw}

\author{Chuan-Chi Wang}
\affiliation{%
  \institution{National Taiwan University}
  \city{Taipei}
  \country{Taiwan}
}
\email{d10922012@ntu.edu.tw}

\author{Chia-Heng Tu}
\affiliation{%
\institution{National Cheng Kung University}
  \city{Tainan}
  \country{Taiwan}
}
\email{chiaheng@ncku.edu.tw}

\author{Shih-Hao Hung}
\affiliation{%
  \institution{National Taiwan University}
  \city{Taipei}
  \country{Taiwan}
}
\email{hungsh@csie.ntu.edu.tw}

\renewcommand{\shortauthors}{B. Trovato et al.}

\begin{abstract}
Quantum approximate optimization algorithm (QAOA) is one of the popular quantum algorithms that are used to solve combinatorial optimization problems via approximations. QAOA is able to be evaluated on both physical and virtual quantum computers simulated by classical computers, with virtual ones being favored for their noise-free feature and availability. Nevertheless, performing QAOA on virtual quantum computers suffers from a slow simulation speed for solving combinatorial optimization problems which require large-scale quantum circuit simulation (QCS). 
In this paper, we propose techniques to accelerate QCS for QAOA
using mathematical optimizations to compress quantum operations, incorporating efficient bitwise operations to further lower the computational complexity, and leveraging different levels of parallelisms from modern multi-core processors,
with a study case to show the effectiveness on solving max-cut problems. 
The experimental results reveal substantial performance improvements, surpassing a state-of-the-art simulator, QuEST, by a factor of 17 on a virtual quantum computer running on a 16-core, 32-thread AMD Ryzen 9 5950X processor. We believe that this work opens up new possibilities for accelerating various QAOA applications.
\end{abstract}

\maketitle
\pagestyle{plain}

\section{Introduction}

Combinatorial optimization searches for an optimal solution from a finite set of feasible solutions, where an objective function is used to calculate and explore the possible solutions under given constraints. Examples of combinatorial optimization problems include maximum cut problem~\cite{maxcut1999}, portfolio optimization~\cite{po1992}, and route planning~\cite{tsp1956}.
Combinatorial optimization has important applications in different fields, such as auction theory, IC design, and software engineering. Many combinatorial optimization problems (e.g., the max-cut problem) cannot be solved in polynomial time. In such a case, it is a common method to resort to approximation algorithms (e.g., genetic algorithm~\cite{genetic}, Monte Carlo simulation~\cite{monte}, and simulated annealing~\cite{sa}) to generate good solutions for such problems in polynomial time.

\begin{figure}[tb!]
\centerline{\includegraphics[scale=0.33]{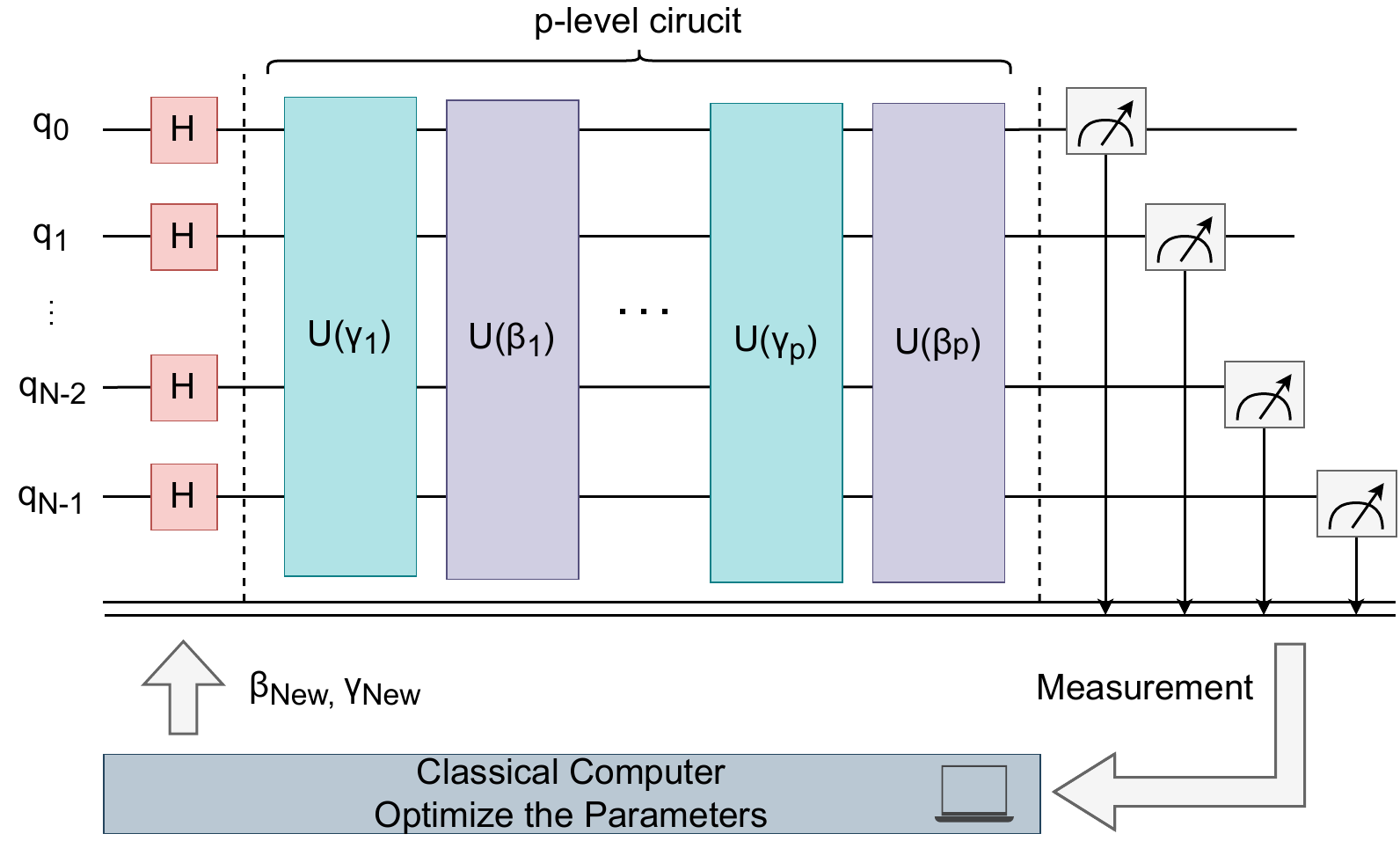}}
\caption{A general architecture of the QAOA circuit for a max-cut problem.}
\label{fig:QAOA_workflow}
\end{figure}

With the advancements in quantum computing, quantum algorithms offer promising alternatives for solving combinatorial optimization problems, where a target problem can be mapped onto a quantum circuit model to be run on a quantum computer to solve the problem. Popular quantum algorithms includes quantum annealing~\cite{QA}, quantum approximate optimization algorithm (QAOA)~\cite{ori_qaoa}, quantum adiabatic algorithm~\cite{QAA}, Shor's algorithm~\cite{shor}, and variational quantum eigensolver~\cite{VQE}. QAOA belongs to the class of quantum-classical algorithms, and this algorithm can exploit classical optimization on quantum operations to tweak an objective function. As illustrated in \figurename~\ref{fig:QAOA_workflow}, a problem is mapped to the quantum circuit (e.g., the \emph{unitary operations}, $U(\gamma)$ and $U(\beta)$) that adjusts the angles of parameterized quantum gate operations to optimize the expected value of the problem, and the relevant parameter optimization is performed on classical computers. As the updating process iterates, the solution is expected to progressively improve.



A QAOA algorithm can be adopted to solve a max-cut problem either on a physical or virtual quantum computer. As a physical quantum computer has a relatively higher cost and is susceptible to noise (leading to errors in the obtained results), a virtual counterpart established by computer simulations is a more preferable alternative, especially for the early stage of the algorithm development because it is free from the noise, if desired. Unfortunately, while a quantum circuit simulator (e.g., QuEST~\cite{QuEST}, Quantum++~\cite{quantum_plus2}, qHiPSTER~\cite{qHiPSTER}, projectQ~\cite{projectQ}, and Cirq~\cite{Cirq}) can serve as the platform to evaluate the QAOA algorithm (e.g., to approximate the result for a max-cut problem), its simulation speed is slow and the simulation time grows almost linearly when a more accurate result is desired (refer to \sectionname~\ref{chap:related_work} for theoretical background and \sectionname~\ref{sec:plevel} for experimental results).

This work aims to accelerate the simulations of the QAOA circuit for solving a max-cut problem. Our proposal is a comprehensive approach as it covers the optimizations on both quantum and classical computing. That is, we propose the mathematical analysis to compress the quantum operations of the standard QAOA algorithm (\sectionname~\ref{sec:general_opt}), adopting the bitwise operations to further lower the computation complexity (\sectionname~\ref{sec:uGraph_opt}), and exploiting different levels of parallelisms to unlash the computing power of multicore/manycore processors (\sectionname~\ref{sec:avx_opt}). The experimental results show that our work outperforms the standard QAOA algorithm running with the state-of-the-art simulator (QuEST~\cite{QuEST}) by a factor of 17.1 and 8.3 on the multicore processor and the manycore processor, respectively. We believe this work can be further generalized to support the acceleration of QAOA algorithms targeting various problems. 

The contributions made by this work are summarized as follows.
\begin{enumerate}
    \item A comprehensive approach is proposed to optimize the QAOA simulation for a max-cut problem from both quantum and classical computing perspectives as described above. To the best of our knowledge, we are unaware of any other work that serves such a purpose with data analysis.

    \item An open-source implementation of the proposed optimizations is online available\footnote{\url{https://anonymous.4open.science/r/QAOA_simulator-8FF3}}
    and there are two versions, one for the x86-based multicore processor and the other for the NVIDIA GPU-based manycore processor.
    
    \item A series of experiments is conducted to demonstrate the efficiency of our proposed optimizations, in terms of simulation speed. Our optimizations achieve a significant improvement of up to two orders of magnitude over the state-of-the-art simulator.
\end{enumerate}


The rest of this paper is organized as follows. \sectionname~\ref{chap:related_work} gives the background and the existing works for the QAOA simulation for a max-cut problem. \sectionname~\ref{chap:method} offers the standard QAOA simulation algorithm and the proposed optimization techniques. \sectionname~\ref{chap:eva} presents the preliminary results. \sectionname~\ref{chap:conclusion} concludes this work.

\section{Methodology}
\label{chap:method}
The QAOA simulation methodologies and the proposed optimization techniques are presented. \sectionname~\ref{sec:qaoasim} elaborates the baseline methodology for the QAOA simulation and introduces the required parameters. \sectionname~\ref{sec:general_opt} presents the developed mathematical optimizations to accelerate the QAOA simulation. \sectionname~\ref{sec:uGraph_opt} offers the optimization that can be used when the max-cut problem can be modelled as an unweighted graph. In addition to the algorithmic optimizations, \sectionname~\ref{sec:avx_opt} introduces the parallelization methodology that utilizes the single-instruction-multiple-data scheme for the developed algorithms to boost simulation performance.


\subsection{The QAOA Simulation} \label{sec:qaoasim}
The QAOA simulation is performed through the state vector-based scheme on the three-layer quantum gates illustrated in \figurename~\ref{fig:circuit}. The important parameters involved with the quantum circuit simulation for QAOA are listed in \tablename~\ref{tab:parameter}. 
The simulation flow, especially for the max-cut problem, is presented in Algorithm~\ref{algo:standard_QAOA}.

Given a quantum system with $N$ qubits, the state vector-based simulation scheme requires $2^N$ complex numbers, each of which represents the amplitude of a specific possible quantum state. The index to the state vector ($stateVector$) is denoted as $b$, where $b$ is usually represented as binary numbers to better correlate to the qubits, and thereby, $b_{i}$ represents the $i$-th bit of $b$ in binary representation. $P$ defines the termination criterion of QAOA, and a larger $P$ value would ordinarily lead to a better approximation ratio. $\gamma_{i}$ and $\beta_{i}$ correspond to the rotation angles and the cost of the mixer layer in $i$-th level, respectively. These angles are essential in modifying the quantum states during the optimization process. The data structure $graph$ is used to represent the node connectivity of the graph for the max-cut problem, and $w$ is the weight of each graph node. 
For instance, $graph_{i,j}$ indicates the connectivity between node$_i$ and node$_j$ of $graph$, and $w_{i,j}$ is the corresponding weight.
The integers $b_I$ and $graph_I$ are used to accelerate the performance of the unweighted graph problem, which will be described later.




\begin{table}[tb!]
  \caption{The parameters required by the QAOA simulation.}
  \label{tab:parameter}
  \begin{tabular}{p{1.3cm}p{6.5cm}}
    \toprule
    Name & Description \\
    \midrule
    $N$ & The total number of qubits simulated \\
    $stateVector$ & The state composed of $2^{N}$ amplitude \\
    $b$ & The index of the $stateVector$  \\

    $P$ & The number of layers in QAOA circuit \\
    $\gamma_i$ & The rotation angle of cost layer in the $i$-th level\\
    $\beta_i$ & The rotation angle of mixer layer in the $i$-th level\\
    $graph$ & The upper triangular matrix used to determine the connectivity of each node\\
    $w$ & The upper triangular matrix used to determine the weight of each node \\
    $b_{I}$ & The integer where each bit is $b_i$\\ 
    $graph_{I}$ & The integer to represent the connectivity of the $i$-th node \\
    \bottomrule
\end{tabular}
\end{table}

Algorithm~\ref{algo:standard_QAOA} provides the pseudocode of the quantum circuit simulation for the single-level QAOA, consisting of three essential layer types, \emph{Initial Layer}, \emph{Cost Layer}, and \emph{Mixer Layer}, as illustrated in \figurename~\ref{fig:circuit}.
The algorithm follows a standard QAOA that starts with Hadamard gates to all qubits for initialization (\emph{initial layer}). It then explores the solutions via applying $p$-levels of updating layers (\emph{$P$-level QAOA}, including cost and mixer layers). Through iterative searching for the required parameters at the updating layers, QAOA is able to deliver a viable approximated solution to the given optimization problem. 

    \begin{algorithm}[tb!]
    \caption{A typical methodology for the QAOA simulation.}
     
    \begin{algorithmic}[1]
    
    \State $stateVector \gets initZeroState()$
    
    \For{$0 < i < N$} \Comment{Initial Layer}
        \State $stateVector \gets HGate(stateVector, i)$
    \EndFor
    
    \For{$0 < p < P$} \Comment{$P$-level QAOA}
    

    \For{$0 \leq i < N$}  \Comment{Cost Layer}
        \For{$i < j < N$} 
            \If {$graph_{i,j}$}
                \State $\theta = w_{i,j}*\gamma_p$
                \State $stateVector \gets RZZGate(stateVector, i, j, \theta)$
            \EndIf
        \EndFor
    \EndFor

    \For{$0 < i < N$} \Comment{Mixer Layer}
        \State $stateVector \gets RXGate(stateVector, i, \beta_p)$
    \EndFor
    
    \EndFor
    \end{algorithmic}
    \label{algo:standard_QAOA}
    \end{algorithm}

\subsection{Mathematical Optimizations}
\label{sec:general_opt}
\begin{figure}[tb!]
\centerline{\includegraphics[width=.9\columnwidth]{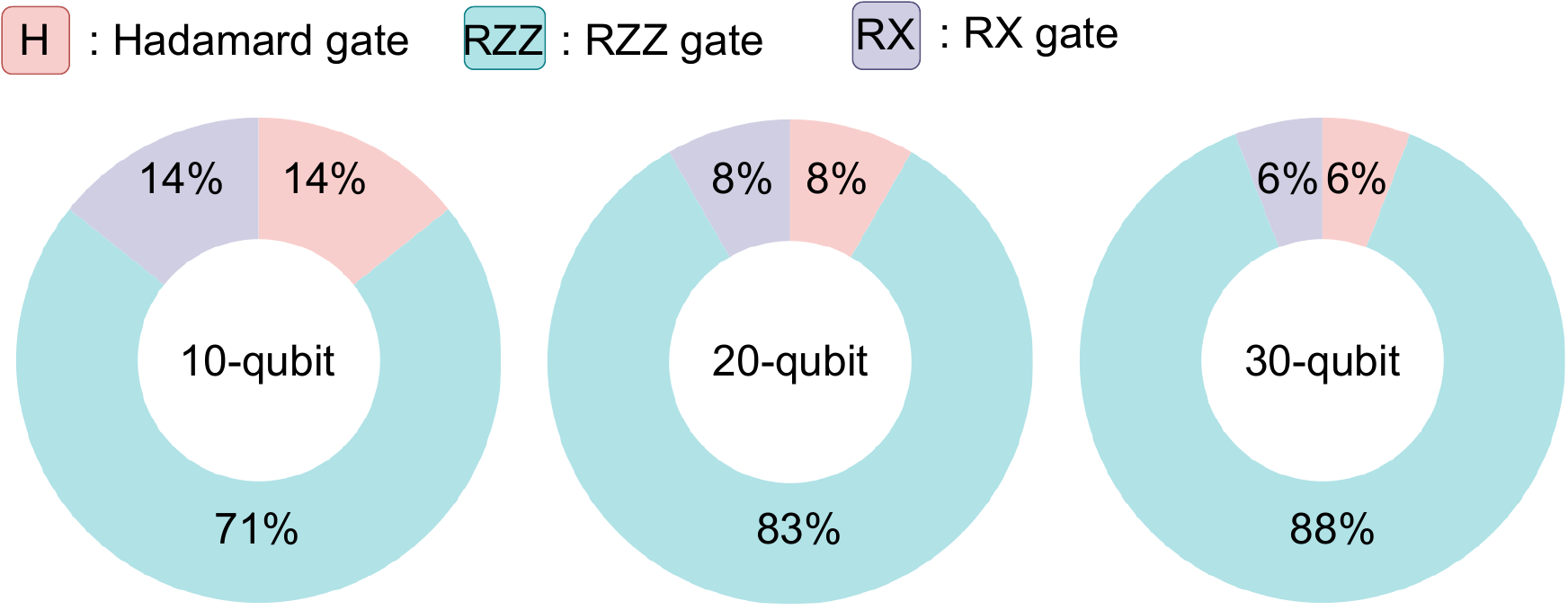}}
\caption{The proportions of the three gate types for a single-level of QAOA.}
\label{fig:circle}
\end{figure}

In a quantum circuit simulation, when the simulation time of one quantum gate is similar to that of another gate, the overall simulation time is proportional to the number of gates to be simulated. In \figurename~\ref{fig:circuit}, it is obvious that the simulation of the RZZ gates dominates the overall performance. The charts shown in \figurename~\ref{fig:circle} display the ratios of the three types of gates when the qubits to be simulated increase from 10 to 30, where the percentages of the RZZ gates grow from 71\% to 88\%. The ratio of the three gate types (H, RZZ, and RX gates) for the single-level QAOA is $2$:$N$-$1$:$2$ (derived from $N$:$\frac{N(N-1))}{2}$:$N$). With the gate ratio in mind, the optimization of RZZ gates becomes the first target for the mathematical optimization, \emph{rotation compression}, which is followed by the optimization for the Hadamard gates, \emph{launch control}.

\paragraph{\textbf{Rotation compression}}
This optimization is driven by the inherent diagonal property demonstrated by RZZ gates.
It can transform the gate-by-gate accesses into state-by-state accesses. As a result, the cache miss ratio during the QAOA simulation can be greatly reduced. This optimization is presented with the following equations, starting from the rewriting of the expression of the cost unitary from Equation~\ref{eq:C} into Equation~\ref{eq:rewrite_UC}.

\begin{equation}
\begin{aligned}
\label{eq:rewrite_UC}
U(\hat{H_C}, \gamma) = e^{-\frac{1}{2} i \gamma { \sum_{\{i,j\} \in E} w_{i,j} Z_i Z_j }}
\end{aligned}
\end{equation}

Furthermore, based on the definition of a Pauli-Z gate ($Z_i = \sum_{b \in \{0, 1\}^N} (-1)^{b_i} \ket{b} \bra{b}$) that represents the unitary as the summation of matrics in the bra-ket notation, Equation~\ref{eq:rewrite_UC} can be derived to Equation~\ref{eq:UC_exp}. 


\begin{equation}
\begin{aligned}
\label{eq:UC_exp}
U(\hat{H}_C, \gamma) & = e^{
  -\frac{1}{2}i\gamma \sum\limits_{\{i,j\} \in E}
  w_{i,j}
  {
    \left (
      \sum\limits_{b \in \{0,1\}^N}(-1)^{b_i}
       \ket{b} \bra{b}
    \right )
    \left (
      \sum\limits_{b \in \{0,1\}^N}(-1)^{b_j}
       \ket{b} \bra{b}
    \right )
  }
}
\end{aligned}
\end{equation}

In accordance with the definition, the inner product of a bra $\bra{b}$ and a ket $\ket{b}$ has the following property.

\begin{equation}
\begin{aligned}
\label{eq:property}
\bra{b} \ket{b'} = 
\begin{cases}
1, & \text{if} \ b=b' \\
0, & \text{otherwise}
\end{cases}
\end{aligned}
\end{equation}

Thanks to the bra-ket property in Equation~\ref{eq:property}, the multiplication is reduced from $2^{2N}$ to $2^{N}$. The rewritten form is shown in Equation~\ref{eq:UC_opt}, where a ket-gra formalism is used to simplify the calculation, and the sign term is intuitively replaced with the XOR operator.


\begin{equation}
\begin{aligned}
\label{eq:UC_opt}
U(\hat{H_C}, \gamma) &= \sum_{b \in \{0,1\}^N} e^{-\frac{1}{2}i\gamma \sum_{\{i,j\} \in E} w_{i,j}{(-1)^{b_i \oplus b_j}}} \ket{b} \bra{b}\\ 
\end{aligned}
\end{equation}

Given the rotation compression optimization in Equation~\ref{eq:UC_opt}, Algorithm~\ref{algo:standard_QAOA} (line 6 $\sim$ 13) can be replaced by Algorithm~\ref{algo:opt_w}, where the outer loop iterates over for updating the $2^N$ amplitudes of the quantum state while the inner loop evaluates multiple RZZ gates (through subtraction or addition operations at line 8 and 10). In order words, thanks to the diagonal property of RZZ gates, the tweaked algorithm performs the rotations of RZZ gates first before the corresponding results are deposited into the $stateVector$ at line 15. On the contrary, the ordinary simulation method shown in Algorithm~\ref{algo:standard_QAOA} requires a state content update (a memory access) for each simulated RZZ gate, incurring a potentially higher number of memory accesses and hence, a higher cache miss rate. 
To be exact, the number of memory accesses required for updating the entire $stateVector$ in Algorithm~\ref{algo:standard_QAOA} is $O(N^2)$, whereas in Algorithm~\ref{algo:opt_w}, it is reduced to $O(1)$.
In summary, this optimization 1) performs consecutive operations on a single quantum state enhancing cache efficiency and 2) converts all multiplications into additions, each quantum state requires only one single rotation.




\begin{algorithm}[bt!]
\caption{The optimized simulation method of the cost layer.}
\label{algo:opt_w}
\begin{algorithmic}[1]
\Function{$rotationCompression$}{$stateVector, graph, w, \gamma$}
    \For{$0 \leq b < 2^N$} \Comment{Scan $2^N$ state}
        \State $ totRotation \gets 0 $
        \For{$0 \leq i < N$}
            \For{$i < j < N$}            
                \If {$graph_{i,j}$}
                    \If {$(b_i \oplus b_j) == 1$}
                        \State $totRotation \gets totRotation - w_{i,j}$
                    \Else
                        \State $totRotation \gets totRotation + w_{i,j}$
                    \EndIf
                \EndIf
            \EndFor
        \EndFor
        \State $stateVector[b] \gets stateVector[b] * e^{-\frac{1}{2}i\gamma \ totRotation}$
    \EndFor
\EndFunction

\end{algorithmic}
\end{algorithm}

\paragraph{\textbf{Launch control}}
In Algorithm~\ref{algo:standard_QAOA}, the Hadamard gates should be applied to all the qubits for the state initialization, where each amplitude is set to $(1 / \sqrt{2})^N$. Based on this observation, the simulation can all $2^N$ amplitudes to $(1 / \sqrt{2})^N$, so as to avoid the simulation of $N$ Hadamard gates. Together with the rotation compression optimization, the simulation speed of the single-level of QAOA can be further reduced. The impact of this optimization is further discussed in \sectionname~\ref{chap:eva}. 



\subsection{Unweighted Graph Optimization}
\label{sec:uGraph_opt}
Algorithm~\ref{algo:opt_w} is developed to deal with the max-cut problem that can be modelled as a weighted graph, where each edge has its own weight/cost value. If the max-cut problem can be modelled as an unweighted graph, where every edge has the same weight/cost of one (i.e., $w_{i,j}=1$), Algorithm~\ref{algo:opt_w} can be further optimized via bitwise instructions rather than addition and subtraction operations. In particular, the rotations can be computed by bitwise operations and the number of `1' bits corresponds to the number of rotations (which can be obtained by a special operation, called population count \textit{popcount}).




The parameters used in this unweighted graph optimization include $totEdge$, $b_I$, and $graph_I$. $totEdge$ is a variable representing the total number of edges in a given graph. $b_I$ is an $N$-bit integer $b_I = (b_i, b_i, ..., b_i)$, where $b_i$ is the $i$-th bit of the integer $b$ repeating $N$ times. 
Similarly, $graph_I$ is also an $N$-bit integer, revealing the connectivity of the $i$-th node of the given graph.
Note that the entries of the $i$-th row of $graph
$ matrix are sequentially arranged into a one-dimensional array in $graph_I$ = ($graph_{i,N-1}$, $graph_{i,N-2}$, ..., $graph_{i,0}$) in binary representation.

Algorithm~\ref{algo:unweight} provides the optimized version to replace the pseudocode (line 5 $\sim$ 13) in Algorithm~\ref{algo:opt_w}. The XOR operation between $b$ and $b_{I}$ is used as a replacement for the XOR operation performed individually on each element in the $i$-th row of $graph$, based on the concept specified in Equation~\ref{eq:UC_opt}. $graph_{I}$ serves as a bitmask for assessing the connectivity of individual nodes while eliminating redundant bits of the leading $i$ bits.
The summation of the `1' bits is accomplished by a \textit{popcount} operation. 
The final result $totRotation$ can be obtained with the arithmetic operations on $totEdge$.

\begin{algorithm}[tb!]
    \caption{The optimized codes used to accumulate the number of cut/uncut edges in the unweighted graph.}
    \label{algo:unweight}
    \begin{algorithmic}[1]
    
            \State $c_{neg} \gets 0$  \Comment{Count the cut edges}
            \For{$0 \leq i < N$}
                \State $b_I \gets \neg((b\verb|>>|i) \land 1) + 1$ \Comment{Expand a bit $b_{i}$ to an integer $b_I$}
                \State $c_{neg} \gets c_{neg} + popCount(graph_I \land (b_I \oplus b ))$
            \EndFor
            \State $c_{pos} \gets totEdge - c_{neg}$  \Comment{Count the uncut edges}
            \State $ totRotation \gets c_{pos} - c_{neg}$
    
    \end{algorithmic}
    \end{algorithm}

\begin{figure}[tb!]
\centerline{\includegraphics[scale=0.35]{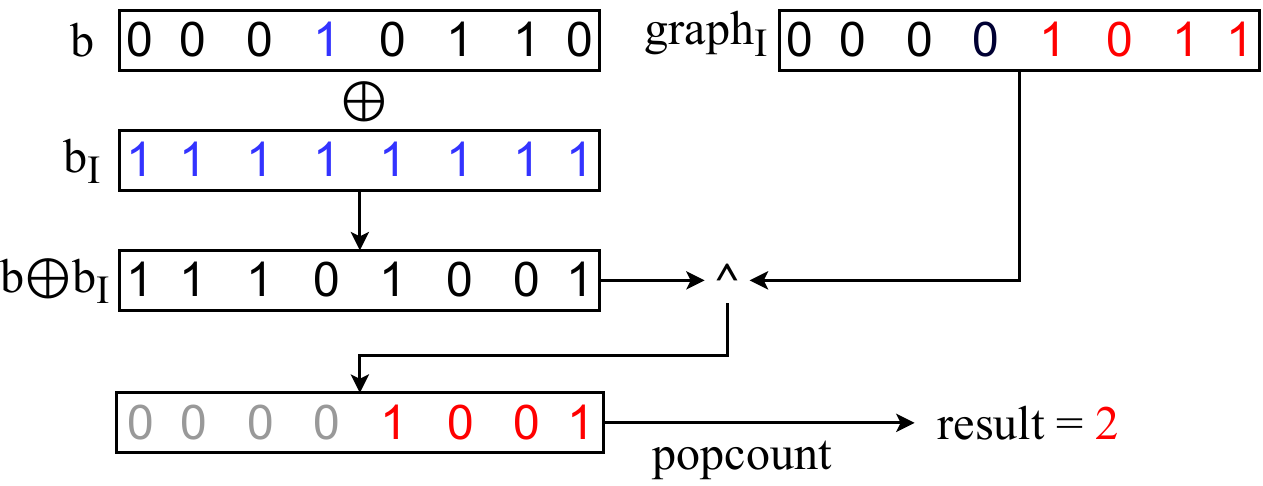}}
\caption{An example of unweighted graph optimization to count the cut edges, where all parameters are 8-bit integers.}
\label{fig:popcount}
\end{figure}

\figurename~\ref{fig:popcount} provides an example of Algorithm~\ref{algo:unweight}, where the given values include an 8-bit integer $b = (00010110)_2$, an 8-bit integer $graph_I = (00001011)_2$, and the indexing value $i = 3$.
After the XOR operation of $b$ and $b_I$,  the resulting value is $(11101001)_2$, which can be further processed by \textit{anding} with $graph_{I}$ to obtain $(00001001)_2$. The \textit{popcount} operation is applied on the result value to obtain the \emph{count}, which is two in this case. By applying these bitwise operations, a loop layer (located at line 5 of Algorithm~\ref{algo:opt_w}) can be removed to lower the computation complexity.



\subsection{Parallel Optimization}
\label{sec:avx_opt}

The QAOA simulation takes advantage of both thread- and instruction-level parallelisms that can be applied on a multicore processor (CPU) and a manycore processor (GPU). The thread-level parallelism exists at the outermost loop (line 2 of Algorithm~\ref{algo:opt_w}). The OpenMP library is used to parallelize the outermost loop, where parallel threads are running to process the loop body of different loop iterations on a CPU. A similar parallel execution is also available on a GPU, which can be done with the CUDA library support. 

The single-instruction-multiple-data (SIMD) support on a processor can take advantage of the instruction-level parallelism. Modern compilers support the auto-vectorization optimization to transform an input code segment into a parallel version utilizing the SIMD instructions of an underlying processor. Nevertheless, the auto-vectorization optimization cannot directly apply to Algorithm~\ref{algo:opt_w} and Algorithm~\ref{algo:unweight} because 1) the conditional statements (i.e., \textit{if-else} statement) within the loops and the unsupported functions (i.e., the exponential function) in Algorithm~\ref{algo:opt_w}, and 2) the unsupported function (i.e., the \textit{popcount} operation) on the experimental hardware processor in Algorithm~\ref{algo:unweight}. Therefore, the algorithms are manually converted into the style accepted by compilers to take advantage of the underlying parallel hardware with the Advanced Vector Extensions (AVX) instruction set support to accelerate the computations. 

The innermost loop is partitioned into sections by using the strip-mining technique, where the length of a strip (or a loop section) is less than or equal to the maximum vector length of the underlying processor. For example, AVX-512 represents a vector operation that can handle a 512-bit number at a time. Furthermore, the loop iteration numbers are defined as \textit{constant} variables, so that compilers can partition the loop iterations at the program compile time. In addition, the statements in the loop body are converted into AVX intrinsic functions to perform the parallel computations. Lastly, the unsupported operations should be replaced with alternative implementations. For example, a lookup table approach is used to handle the \textit{popcount} operations~\cite{avx_popcnt}. The performance results of the SIMD optimization are provided in \sectionname~\ref{sec:exp_cpu}. 

\section{Evaluation}
\label{chap:eva}
The experimental environment is introduced in \sectionname~\ref{sec:setup}. The performance improvements achieved by the multicore processor and the manycore processor (using the optimizations described in \sectionname~\ref{chap:method})are presented in Sections~\ref{sec:exp_cpu} and ~\ref{sec:exp_gpu}, respectively. These performance data are compared against the state-of-the-art work, QuEST. The quality of the approximation results under different $p$ values ($p$-level QAOA circuits) is discussed in \sectionname~\ref{sec:plevel}.


\subsection{Experimental Setup}
\label{sec:setup}
The hardware and software configurations that are used for the QAOA simulation are listed in \tablename~\ref{tab:platform}. The multicore processor (AMD Ryzen 9) and the manycore processor (NVIDIA RTX 4090) are used for the QAOA simulation, and the delivered performance results are compared against those data obtained by the state-of-the-art simulator QuEST~\cite{QuEST}, referred to as \emph{baseline} that implements the pseudocode in Algorithm~\ref{algo:standard_QAOA}. Different experimental configurations are represented by the variables, $Opt$ for the adoption of an optimization technique introduced in Section~\ref{chap:method} and $SU$ for the speedup compared to the baseline. Specifically, $Opt_{w}$ and $Opt_{u}$ denote the optimizations for the weighted and unweighted graph, respectively, where the optimizations are run with parallel threads. Furthermore, $Opt_{uAVX}$ means that the AVX instructions are further used for the computation acceleration, as introduced in \sectionname~\ref{sec:avx_opt}. By default, a five-level QAOA is used as the target quantum circuit for our QAOA simulation and for the QuEST simulation. The five-level QAOA is chosen since it would provide superior approximation ratios than traditional approximation algorithms for the benchmark of the u3R and w3R graph~\cite{level_numbers}.


\begin{table} [tb!]
\caption{The hardware and software platforms.} 
  \label{tab:platform}
  \small{
  \begin{tabular}{p{1.3cm}p{6.5cm}}
    \toprule
   \textbf{Name} & \textbf{Attribute} \\
    \midrule
    CPU & AMD Ryzen 9 5950X Processor (w/ 32 threads \& AVX2)\\
    GPU & NVIDIA GeForce RTX 4090 (16,3884 cores, 24 GB mem.)\\
    RAM & Kingston 128 GB (=4 * 32GB) DDR4 2400MHz \\ 
    \midrule
    OS  & Ubuntu 22.04 LTS (kernel version 5.19.0-43-generic)\\
    CUDA & CUDA Toolkit version 12.1.105 \\
    Compiler & GCC version 12.2.1 \\
    OpenMP & OpenMP version 4.5 \\
    \bottomrule
\end{tabular}
  } 
\end{table}

\subsection{Multicore Processor}
\label{sec:exp_cpu}
\begin{figure}[tb!]
\centerline{\includegraphics[scale=0.4]{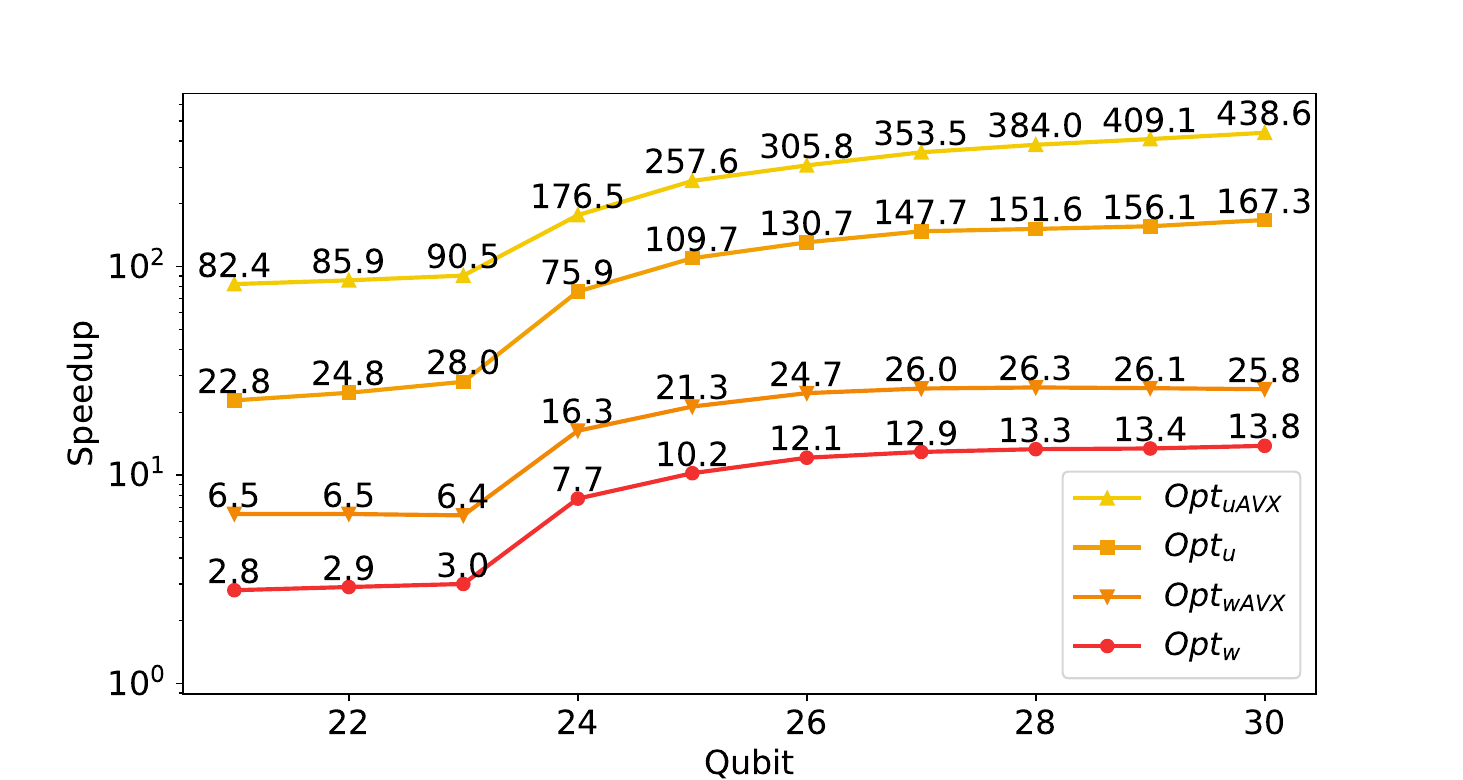}}
\caption{The speedup of the cost layer on the multicore processor.}
\label{fig:cpu_speedup}
\end{figure}

The five-level QAOA is used as the input to QuEST and our simulator with the qubit size ranging from 21 to 30, where the state vectors are kept in 128 GB of the main memory. \tablename~\ref{tab:cpu} lists the overall performance data delivered by QuEST with parallel threads (\emph{Baseline}), our simulator with parallel threads ($Opt_{w}$), and with parallel threads plus AVX support ($Opt_{wAVX}$) for a weighted graph. In order to observe the performance improvements achieved by the proposed optimizations elaborated in Sections~\ref{sec:general_opt} and~\ref{sec:uGraph_opt} on the cost layer, \figurename~\ref{fig:cpu_speedup} plots their speedups to show the performance trends of the optimizations under a different number of qubits. 

\paragraph{\textbf{Overall performance}} The speedups $SU_{w}$ and $SU_{u}$ in \tablename~\ref{tab:cpu} indicate that the optimized Algorithm~\ref{algo:opt_w} and Algorithm~\ref{algo:unweight} can reduce the computational complexity (turning the rotations into additions/subtractions in Algorithm~\ref{algo:opt_w} and using the bitwise operations in Algorithm~\ref{algo:unweight}), resulting in up to 7.5x and 14.2x speedups against the parallel version of QuEST (\emph{Baseline}), respectively. The AVX support can further improve the performance by a factor of 2.1 (for $SU_{wAVX}$) and 2.7 (for $SU_{uAVX}$) on average. It is important to note that when the qubit size is larger than 24, the quantum states cannot be accommodated by the L3 cache of the processor (64 MB). In this case, the vectorized bitwise operations slightly outweigh the overhead of the data movements to pack and unpack the data, and the performance improvements achieved by the AVX feature ($SU_{uAVX}$) drop from 3.1x (23 qubits) to 0.2x (24 qubits).

\begin{table*}[bt!]
    \caption{The elapsed time of 5-level QAOA on the multicore processor with different optimizations (unit: milliseconds).}
    \label{tab:cpu}
    \setlength{\tabcolsep}{3.8mm}{
    \begin{tabular}{crrrrrrrrr}
    \hline \specialrule{0em}{1.5pt}{1.5pt}
    Qubit & $Baseline$ & $Opt_{w}$ & $SU_{w}$ & $Opt_{wAVX}$ & $SU_{wAVX}$ & $Opt_{u}$ & $SU_{u}$ & $Opt_{uAVX}$ & $SU_{uAVX}$ \\
    \hline
        21 &       325 &     124 & 2.6x &      63 & \textbf{5.2x} (+2.6) &     30 & 10.9x &     21 & \textbf{15.5x} (+4.6) \\
        22 &       710 &     264 & 2.7x &     135 & \textbf{5.2x} (+2.5) &     61 & 11.6x &     42 & \textbf{17.1x} (+5.5) \\
        23 &     1,940 &     575 & 3.4x &     308 & \textbf{6.3x} (+2.9) &    145 & 13.4x &    118 & \textbf{16.5x} (+3.1) \\
        24 &     8,418 &   1,762 & 4.8x &   1,164 & \textbf{7.2x} (+2.4) &    798 & 10.5x &    790 & \textbf{10.7x} (+0.2) \\
        25 &    24,333 &   4,159 & 5.8x &   3,020 & \textbf{8.1x} (+2.3) &  2,100 & 11.6x &  1,967 & \textbf{12.4x} (+0.8) \\
        26 &    60,794 &   9,257 & 6.6x &   6,856 & \textbf{8.9x} (+2.3) &  4,962 & 12.3x &  4,693 & \textbf{13.0x} (+0.7) \\
        27 &   138,475 &  20,137 & 6.9x &  15,126 & \textbf{9.2x} (+2.3) & 10,957 & 12.6x & 10,410 & \textbf{13.3x} (+0.7) \\
        28 &   302,501 &  42,128 & 7.2x &  31,898 & \textbf{9.5x} (+2.3) & 22,885 & 13.2x & 21,653 & \textbf{14.0x} (+0.8) \\
        29 &   647,955 &  88,188 & 7.3x &  66,667 & \textbf{9.7x} (+2.4) & 47,221 & 13.7x & 44,724 & \textbf{14.5x} (+0.8) \\
        30 & 1,385,228 & 185,209 & 7.5x & 139,663 & \textbf{9.9x} (+2.4) & 97,482 & 14.2x & 92,291 & \textbf{15.0x} (+0.8) \\
    \hline
    \end{tabular}
    }
\end{table*}


\paragraph{\textbf{Speedups of the cost layer}} The performance enhancements achieved by the cost layer are plotted in \figurename~\ref{fig:cpu_speedup} to observe the performance-dominant layer behaviors under different qubit sizes. It is obvious that the optimizations presented in Sections~\ref{sec:uGraph_opt} and~\ref{sec:avx_opt} have a significant impact on the simulation performance. Furthermore, the performance improvements achieved by the AVX feature are significant, where the ideal performance speedup contributed by the AVX feature is four for using four 64-bit integers to handle the four concurrent operations at a time. It is worth noting that if the max-cut problem can be modelled as an unweighted graph, the speed of the QAOA simulation delivered by our tool is 400x faster than that of the state-of-the-art simulator for the cost layer. 



\subsection{Manycore Processor}
\label{sec:exp_gpu}
The performance speedups $SU_{w}$ and $SU_{u}$ achieved by the GPU are presented in \tablename~\ref{tab:gpu}. It is important to note that the $SU_{w}$ achieved by the GPU is similar to that delivered by the CPU, and this demonstrates that our proposed mathematical optimizations in \sectionname~\ref{sec:general_opt} have substantial performance improvements across different computer architectures. On the contrary, the $SU_{u}$ in \tablename~\ref{tab:cpu} is larger than the $SU_{u}$ in \tablename~\ref{tab:gpu}, which reflects the architectural difference between CPU and GPU. 
Note that the \textit{popcount} operation is implemented by the GCC library for the CPU version and by the CUDA library for the GPU version. 

\begin{figure}[tb!]
\centerline{\includegraphics[scale=0.4]{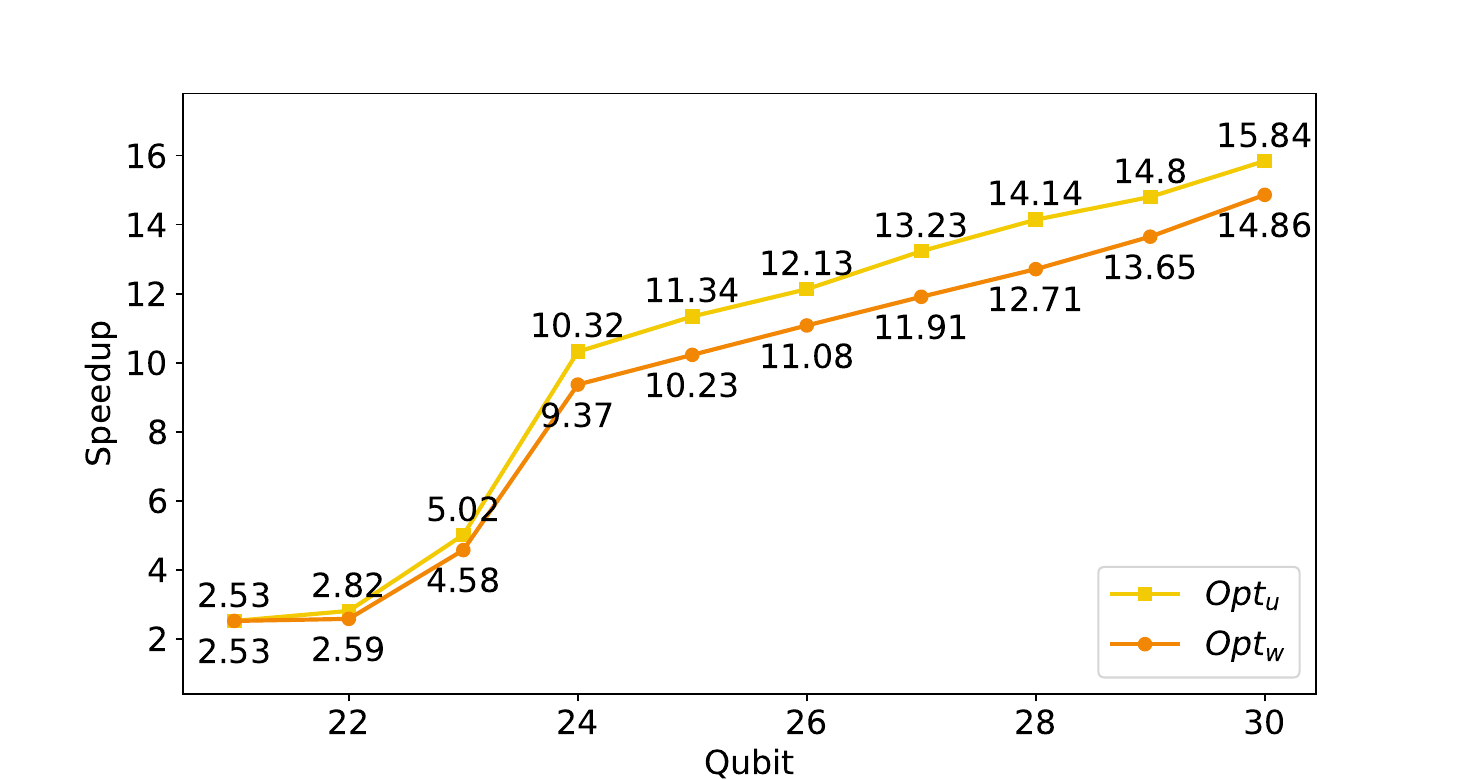}}
\caption{The speedup of the cost layer on the manycore processor.}
\label{fig:gpu_speedup}
\end{figure}

Considering the speedups of the cost layer for the GPU, the simulation time can achieve up to 9.3x speedup ($SU_{w}$) for the 24-qubit configuration against Baseline, and the unweighted graph configuration can have an additional 10\% improvement ($SU_{u}$). The detailed performance data are plotted in \figurename~\ref{fig:gpu_speedup}. 
It is important to note that the GPU simulation time (e.g., $Opt_{w}$ and $Opt_{u}$ in \tablename~\ref{tab:gpu}) is up to two orders of magnitude less than the CPU time (in \tablename~\ref{tab:cpu}). It shows that our proposed methods can provide a better performance boost when a GPU is available. Also note that when the qubit size is larger than 23, the simulation time increases significantly because the L2 cache size of the GPU (72 MB) cannot keep all the states. 


\begin{table}[bt!]
    \caption{The elapsed time of 5-level QAOA on the manycore processor with different optimizations (unit: milliseconds).}
    \label{tab:gpu}
    \setlength{\tabcolsep}{3.2mm}{
    \begin{tabular}{crrrrrr}
    \hline \specialrule{0em}{1.5pt}{1.5pt}
    Qubit & $Baseline$ & $Opt_{w}$ & $SU_{w}$ &$Opt_{u}$ & $SU_{u}$ \\
    \hline
        21 &     17 &     8 & 2.1x &     7 & 2.4x \\
        22 &     34 &    17 & 2.0x &    14 & 2.4x \\
        23 &    121 &    38 & 3.1x &    31 & 3.9x \\
        24 &    488 &    96 & 5.1x &    81 & 6.0x \\
        25 &  1,050 &   196 & 5.4x &   168 & 6.2x \\
        26 &  2,247 &   398 & 5.7x &   331 & 6.8x \\
        27 &  4,809 &   811 & 5.9x &   676 & 7.1x \\
        28 & 10,268 & 1,676 & 6.1x & 1,383 & 7.4x \\
        29 & 21,877 & 3,407 & 6.4x & 2,790 & 7.8x \\
        30 & 46,548 & 6,957 & 6.7x & 5,640 & 8.3x \\
    \hline
    \end{tabular}
    }
\end{table}


\subsection{Approximation Result Quality}
\label{sec:plevel}
\begin{figure}[bt!]
\centerline{\includegraphics[scale=0.4]{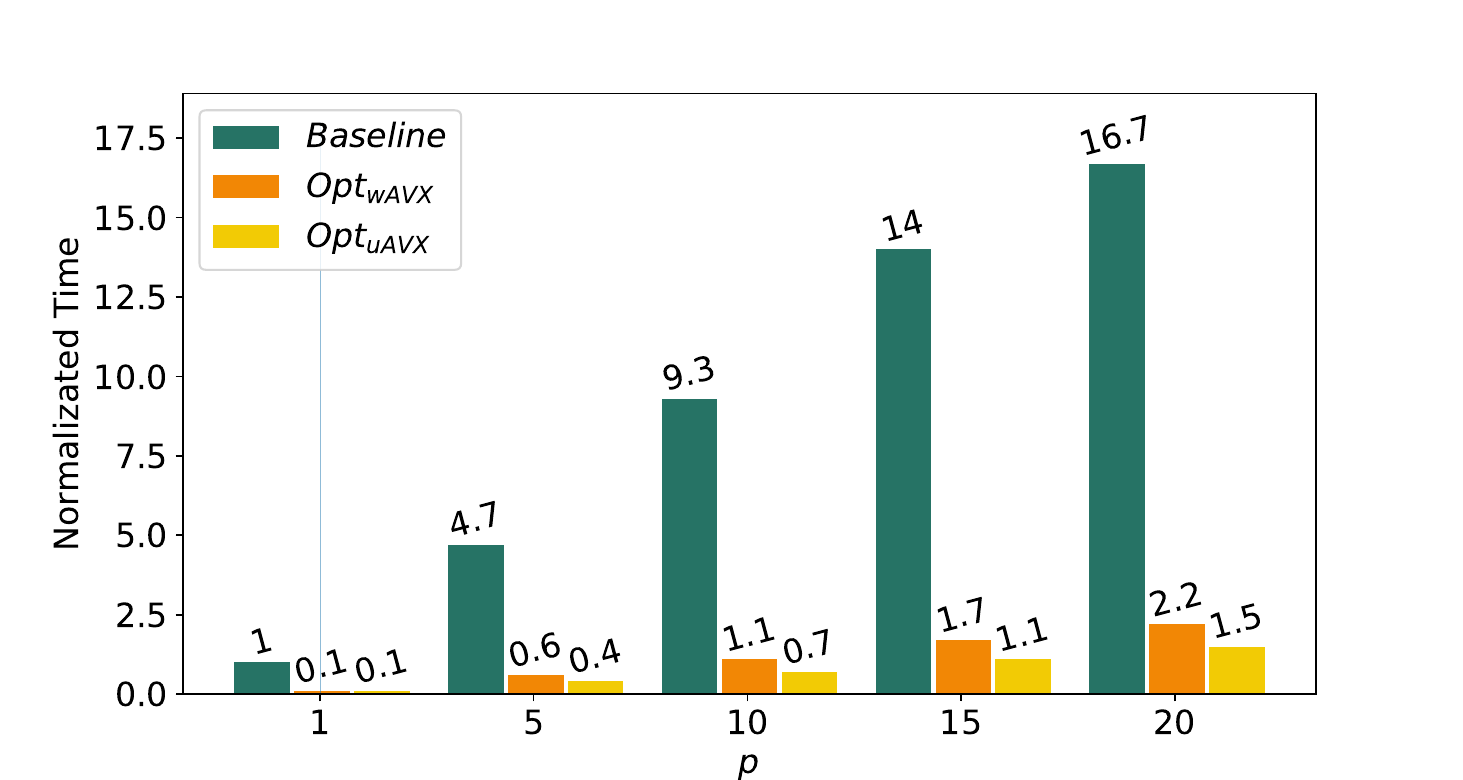}}
\caption{The CPU simulation times vary with different the values of $p$ for three distinct techniques.}
\label{fig:cpu_bar}
\end{figure}

\begin{figure}[bt!]
\centerline{\includegraphics[scale=0.4]{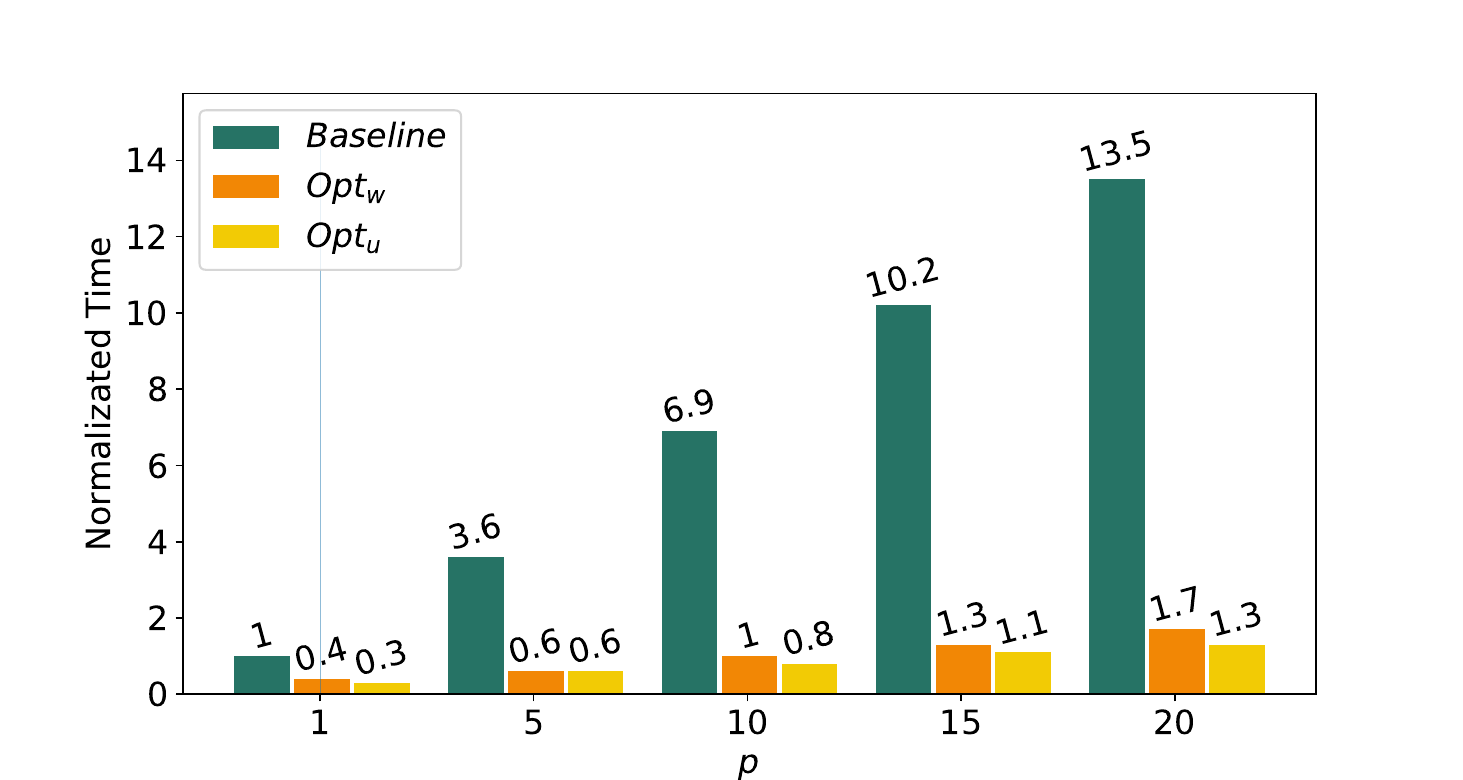}}
\caption{The GPU simulation times vary with different the values of $p$ for three distinct techniques.}
\label{fig:gpu_bar}
\end{figure}

The level of QAOA circuits $p$ relates to the quality of the approximation result as introduced in \sectionname~\ref{sec:qaoacircuit}, and a larger $p$ value leads to a longer simulation time. If an application developer desires a better result, she/he should wait for a longer time. To analyze how the simulation time scales with the increasing of the $p$ value, a 30-qubit circuit is constructed with the $p$ value ranging from 1 to 20. Figs.~\ref{fig:cpu_bar} and~\ref{fig:gpu_bar} plot the normalized time (normalized to the baseline with $p=1$) when the simulations are performed on the CPU and the GPU, respectively.  

Both figures show that the simulation time grows almost linearly with the increasing $p$ value (e.g., approaching the slope of one for the CPU simulation time). On the contrary, our proposed optimization approaches have a relatively shorter simulation time. With ultra-fast simulation speed, our methods are able to deliver far better approximation results (e.g., $p=15$), but only require 20\% more time than the $p=1$ configuration running with the state-of-the-art simulator on the multicore processor. These results evidently indicate that our methods are able to deliver a higher-quality result at a lower cost (simulation time). 





\section{Conclusion} \label{chap:conclusion}
This work aims to boost the performance of the QAOA simulations for the max-cut problem. Different levels of optimizations have been proposed for the simulation acceleration, in terms of mathematical forms (\sectionname~\ref{sec:general_opt}), bitwise implementation scheme (\sectionname~\ref{sec:uGraph_opt}), and parallelization techniques (\sectionname~\ref{sec:avx_opt}). Compared with the state-of-the-art simulator QuEST, our work achieves up to 438x speedups for the cost layer of a single-level QAOA circuit and up to 17x speedups for the five-level QAOA circuit on the multicore processor. A significant performance acceleration is also observed on the manycore processor. We believe that our work paves the way toward efficient QAOA simulations for the max-cut problem.

\bibliographystyle{ACM-Reference-Format}
\bibliography{mybib} 

\end{document}